\documentclass[prl,twocolumn,amssymb]{revtex4}

\usepackage{graphicx}

\begin{document}
\title{Reflections Function Method In the X~-~Ray Reflectometry}
\author{ N.V. Bagrets, E.A. Kravtsov, V.V. Ustinov. }
\address{Institute of Metal Physics, 18, Kovalevskaya St., Ekaterinburg, 620219, 
Russia}
\date{\today}

\begin{abstract}
The theory of  specular X-ray reflectivity from a rough interface based upon
the reflection function method (RFM) is proposed. The RFM transforms the second
order differential equation for the wave amplitude into the non-linear first
order differential equation of Riccati type for the reflection function. 
This equation is solved in the approximation of the abruptly changing 
potential, which is justified for the typical angles of X-ray reflectometry. 
The reflectivity is represented as a series.  The first term of this series 
reproduces the Nevot-Croce approximation and second one gives the phase 
correction for greater angles. It is shown that the phase correction can be 
used to obtain the degree of interface asymmetry. The X-ray reflectometry model 
profiles for Fe/Cr superlattice are used to illustrate the method.

\end{abstract}
\maketitle


The X-ray reflectometry is a useful tool for studying surface and 
interface structure in thin films and multilayers. Usually, the rough surface 
X-ray reflection is analyzed in the framework of the plane-wave Born approximation 
(PWBA) or the distorted-wave Born approximation (DWBA) \cite{Holy,Zhou}. In 
this work, we apply the reflection function method (RFM) \cite{Babikov} to the 
specular X-ray reflection from a rough surface or interface.

Let us consider the X-ray reflection on a non-ideal interface structure. We 
assume that this structure is homogeneous along the surface which is parallel to 
the $(x,y)$ plane and the media can be characterized by its dielectric 
susceptibility $\chi(z)$ depending only on the normal coordinate $z$, where 
$\chi(z)\to 0$ when $z\to\pm\infty$. The change of the material occurs only in 
the $z$-direction perpendicular to the surface. Then one has to solve the 
one~-~dimensional Helmholtz equation
\begin{equation}
\left(\frac{d^2}{dz^2}+k^2\sin^2\theta\right)E(z)+k^2\chi(z)E(z)=0
\label{Helm}
\end{equation}

Here $E(z)$ is the electric field in the medium, $\theta$ is the incident angle 
and $k=2\pi/\lambda$, $\lambda$ being a wave length of radiation. As the first 
step, we need to evaluate the scattering matrix 
$S_{12}=\left(
\begin{array}{cc}
r_{11} & t_{12} \\
t_{21} & r_{22}
\end{array}
\right)
$
related with the given interface between two subsequent layers, which are 
denoted as 1 and 2.
The RFM starts from the transformation of the linear second order 
differential equation (\ref{Helm}) for the wave amplitude $E(z)$ into a non-
linear first order equation of Riccati type for the reflection function $B(z)$. 
This transformation is not unique and can be performed in a number of different 
ways. An advantage of the RFM is that the perturbation expansion carried out in 
the framework of this scheme gives more rapid convergence in comparison with 
the conventional Born series. In particular, the first order approximation 
easily enables one to go beyond DWBA. 

  We denote $q(z)=2k\sqrt{\sin^2\theta+\chi(z)}$, and represent the electric 
field $E(z)$ in the form
\begin{eqnarray}
E(z)=q^{-1/2}(z)\left[ 
A(z)\exp\left( \frac{i}{2}\int_{z_0}^{z}q(x)dx \right) + \right. \nonumber\\
\left. C(z)\exp\left(-\frac{i}{2}\int_{z_0}^{z}q(x)dx \right)
\right], 
\label{E}
\end{eqnarray}
where $A(z)$ and $C(z)$  are amplitude functions. In addition, we apply
the following condition:
\begin{eqnarray}
\frac{d}{dz}E(z)=\frac{i}{2}q^{1/2}(z)\left[ 
A(z)\exp\left( \frac{i}{2}\int_{z_0}^{z}q(x)dx \right)  \right. \nonumber\\
\left. -C(z)\exp\left(-\frac{i}{2}\int_{z_0}^{z}q(x)dx \right)
\right], 
\label{dE}
\end{eqnarray}
The reflection function $B(z)$ is defined as $B(z)=C(z)/A(z)$. Taking into 
account 
the continuity of $E(z)$ and Eq.~(\ref{Helm}) one can prove that $B(z)$ satisfies 
the first order nonlinear differential equation 
\begin{eqnarray}
\frac{d}{dz}B(z)=\frac{q'(z)}{2q(z)}\left[ 
\exp\left( i\int_{z_0}^{z}q(x)dx \right)  \right. \nonumber\\
\left. -B^2(z)\exp\left(-i\int_{z_0}^{z}q(x)dx \right)
\right], 
\label{Ric}
\end{eqnarray}
Eqs. (\ref{Helm}) and (\ref{Ric}) should be supplemented by the boundary 
conditions. For example, the choice of $B(+\infty)=0$ corresponds to the X-ray 
beam, incident from $z<0$, and in this case a reflection coefficient $r_{11}$ is 
given by relation $r_{11}=B(-\infty)$. 
We also introduce into consideration the dimensionless functions 
$g_{\pm}(z)$, which are related to $\chi(z)$ via equality 
$\chi(z)-\chi_\pm=\pm(\chi_- -\chi_+)g_\pm(z)$. The function $g_-(z)\to 0$, when 
$z\to -\infty$, and $g_-(z)\to 1$, if $z\to +\infty$ (See Fig.1). The functions 
$g_{\pm}(z)$ obey the relation $g_+(z)+g_-(z)=1$. One can regard $g_{\pm}(z)$ 
as a "shape" of the interface, which reproduces the gradual transition from the 
first layer to the second one. We shall call interface "symmetric", if 
$(\frac{\partial}{\partial z})g_-(z)$ is an even function of $z$, otherwise 
interface is "asymmetric".

In case of grazing incidence angles Eq.(\ref{Ric}) can be solved in the 
approximation of the abruptly changing potential. The small parameter $\epsilon$ 
of this expansion is defined as $\epsilon=a q_c/2\pi$, where $a$ is 
characteristic length corresponding to the variation of the potential and 
$q_c=max|q(z)|$. In the X-ray reflectometry studies $a$ is of the order of mean-
root-square interfacial roughness $\sigma=2-8$~\AA (See~\cite{Bai}) and $q_c\sim 
(4\pi/\lambda)\sin\theta$. Therefore the condition $\epsilon\le 1$ holds over 
the 
scattering angle region $(0<\theta<4^0)$. These estimates make it possible to 
find the solution of Eq. (3) in the form
\begin{equation}
B(z)=B_0(z)\exp(\beta(z)),
\label{Series}
\end{equation}
where
\begin{eqnarray}
B_0(z)&=&(q(z)-q_2)/(q(z)+q_2), \nonumber\\
q_{2(1)}&=&2k\sqrt{\sin^2 \theta +\chi_{\pm}}, \nonumber
\end{eqnarray}
and
$$
\beta(z)=\sum_{n=1}^{+\infty}\beta_n(z)\epsilon^n
$$

The function $B_0(z)$ corresponds to the boundary condition $B(+\infty)=0$ and 
it gives the Fresnel reflection coefficient $r_{11}=(q_1-q_2)/(q_1+q_2)$ from 
an ideal sharp interface. The series $\beta(z)$ yields the corrections due to 
the interfacial non-ideality.

The use of ansatz (\ref{Series}) is the essential step in the derivation. 
It enables us partially to sum up the reducible parts of the expansion $B(z)$ 
in powers of $\epsilon$, so that the coefficients $\beta_n(z)$ are associated 
with the irreducible terms only. The series $\beta(z)$ can be found by means of 
subsequent iterations from Eq.~(\ref{Ric}). It turns out that, at each step $n$, 
one encounters the only linear inhomogeneous differential equation for 
$\beta_n(z)$. The details of this derivation will be presented elsewhere. As a 
result, up to the third order of $\epsilon$, the elements of the matrix 
$S_{12}$ can be written in the form
\begin{eqnarray}
r_{11} &=& r_{11}^F\exp\left(
i q_1\delta -\frac12 q_1 q_2 \sigma^2 \nonumber \right. \\
&+& \left.  i q_1\left[ (q_1^2 + 3 q_2^2)\mu_1^3 + 
(q_1^2 -q_2^2)\mu_2^3 \right]\sigma^3 \right)
\label{r11}
\end{eqnarray}
\begin{eqnarray}
r_{22} &=& r_{22}^F\exp\left(-iq_2\delta-\frac12 q_1 q_2 \sigma^2 
\right. \nonumber\\
&+&\left. i q_2\left[ -(q_2^2 + 3 q_1^2)\mu_1^3 - (q_1^2 -
q_2^2)\mu_2^3 \right]\sigma^3 \right)
\label{r22}
\end{eqnarray}
\begin{eqnarray}
\label{t12} 
t_{12(21)} &=& t_{12(21)}^F\exp\left(
\frac{1}{2}i(q_1-q_2)\delta+
\frac18 (q_1 - q_2)^2 \sigma^2  \right. \\
&+&\left.\frac 12 i (q_1-q_2)^2 
\left[ (q_1 - q_2)\mu_1^3
+ (q_1 + q_2) \mu_2^3 \right]\sigma^3 \right)
\nonumber
\end{eqnarray}
Here $r^F$, $t^F$ are Fresnel's reflection and transmission amplitudes and 
parameters $\delta$, $\sigma$, and $\mu_{1(2)}$ are expressed via $g_{\pm}(z)$ 
as follows
\begin{eqnarray}
\delta &=& \int_{-\infty}^{+\infty} z\frac{d}{dz} g_{-}(z)\,dz \\
\sigma^2 &=& 2I^{(2)}(-,+) \nonumber\\
&=&2\int_{-\infty}^{+\infty} g_{-}(z_1)dz_1 \label{rgh}
\int_{z_1}^{+\infty}g_{+}(z_2)dz_2   \\
\mu_{1(2)}^3&=&\left[I^{(3)}(-,+,+)\mp I^{(3)}(-,-,+) \right]/4\sigma^3
\end{eqnarray}
where
\begin{eqnarray}
I^{(3)}(-,\pm,+)&=&\int_{-\infty}^{+\infty} g_{-}(z_1)dz_1
\int_{z_1}^{+\infty}g_{\pm}(z_2)dz_2 \nonumber\\
&&\int_{z_2}^{+\infty}g_{+}(z_3)dz_3  \nonumber
\end{eqnarray}

\begin{figure}[t]
\includegraphics[scale=0.4]{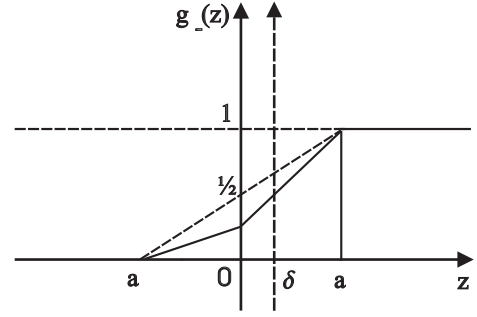}
\caption{The linear segment form of the profile $g_{-}(z)$, corresponding to the 
interface of a width  $2a$. The "symmetric" case is shown by the dashed line, 
and the "asymmetric" one is depicted by the solid line.}
\end{figure}

Consider now the physical meaning of Eqs.~(\ref{r11}-\ref{t12}). First of all, 
the phase shift $\delta$ arises due to transmitting electromagnetic wave in the non-uniform 
interface region. This phase shift is equivalent to "effective" increasing in 
the thickness of layer 1 to value $\delta$: $z=z'-\delta$. (See Fig.1) In the 
process of the numerical treatment of the X-ray reflectometry profiles this 
fact enables one to adjust the ratio between the layers' thickness' in the 
periodical cell of the superlattice in order to obtain the best fit to 
experimental data. The second order correction to the amplitudes $r^F$, $t^F$ 
in Eq. (6-8) reproduces the well-known Nevot-Croce~\cite{Nevot} approximation. 
The magnitude $\sigma$ has the meaning of the root-mean-square interfacial 
roughness and it is given by Eq.(\ref{rgh}).
	
The phase correction corresponding to the third order terms of the expansion is 
a new feature in the question. In addition to $\sigma$ it contains two extra 
parameters $\mu_1$ and $\mu_2$. We found, that $\mu_2$ is in general nonzero 
for a wide set of profiles $\chi(z)$ whereas $\mu_1$ does not vanish in case of 
the asymmetric interfaces only. Hence, this property may be used to define 
$\mu_1$ as  the measure of the interface asymmetry. The parameter $\mu_2$ has no
such an evident meaning as $\mu_1$. But we may note, that if necessary,
it can be eliminated from the consideration. It follows directly from the
structure of Exp.~(6-8). Since the factor 
$q_1^2-q_2^2 = 16(\pi^2/\lambda^2)(\chi_{-}-\chi_+)$ does not depend on $\theta$,
the terms, proportional to $\mu_2^3$, give mere the additive contribution to $\delta$.
One may take it into account by omiting the terms with $\mu_2^3$ and 
substituting $\delta\rightarrow \delta + 16(\pi^2/\lambda^2)(\chi_{-}-\chi_+)\mu_2^3$.

\begin{figure}[t]
\includegraphics[scale=1.0]{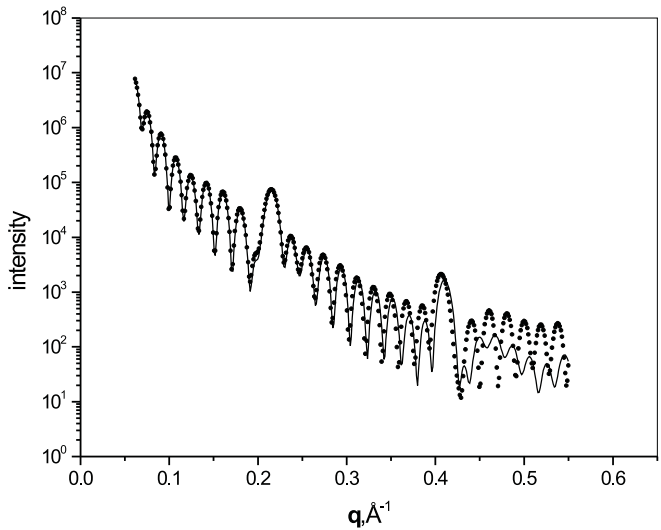}
\caption{Model X-ray reflectivity profiles for multilayer structure 
 Al$_2$O$_3$/Cr(70\AA)/[Fe(20\AA)/Cr(9\AA)]$_8$ calculated without asymmetric
 phase corrections (points), and with asymmetric phase correction 
($\mu_1^3=0.2$, solid line). Wave length  $\lambda = 1.789$\AA.}
\end{figure}
\begin{figure}[b]
\includegraphics[scale=0.42]{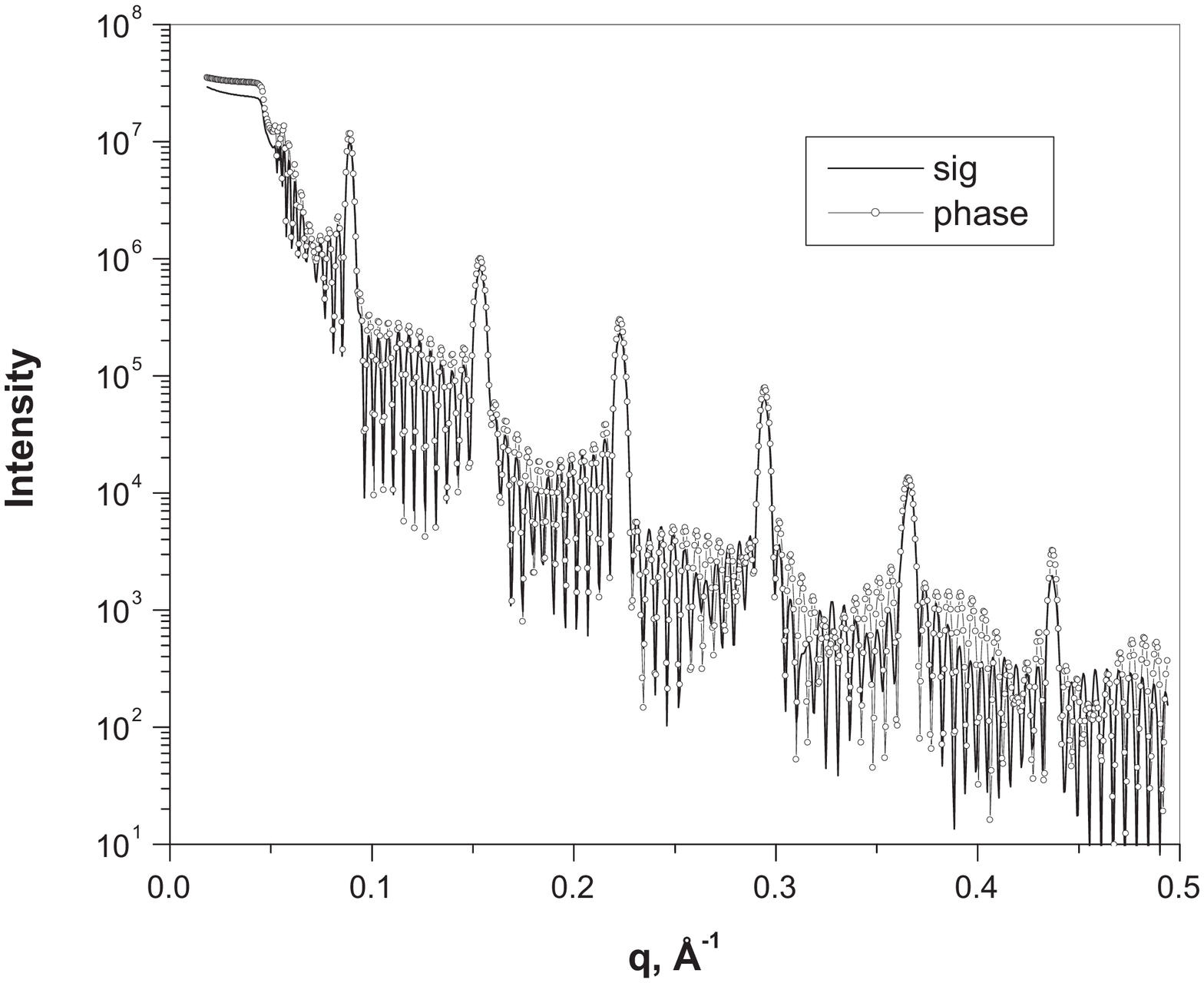}
\caption{Model X-ray reflectivity profiles for multilayer structure 
 Al$_2$O$_3$/Cr(70\AA)/[Fe(70\AA)/Cr(9\AA)]$_{12}$ calculated without asymmetric
 phase corrections (sig), and with asymmetric phase correction 
(phase). Wave length  $\lambda = 2.070$\AA.
}
\end{figure}
\begin{figure}[b]
\includegraphics[scale=0.45]{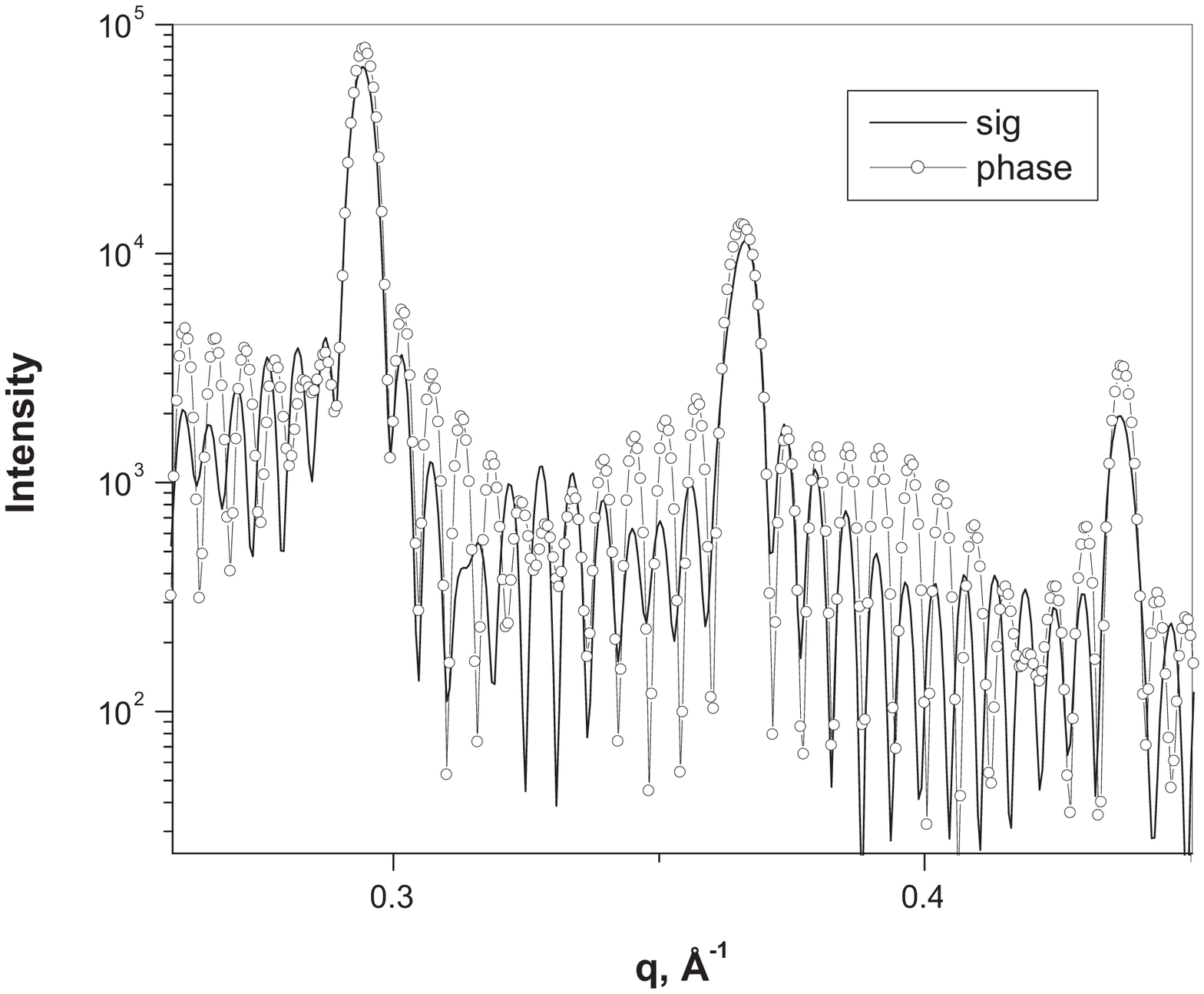}
\caption{Model X-ray reflectivity profiles for multilayer structure 
 Al$_2$O$_3$/Cr(70\AA)/[Fe(70\AA)/Cr(9\AA)]$_{12}$ calculated without asymmetric
 phase corrections (sig), and with asymmetric phase correction 
(phase). Wave length  $\lambda = 2.070$\AA.}
\end{figure}

To test the obtained approximation we exploited the symmetric Epstein 
profile $g_{-}^E(z)=(1+e^{-z/a})^{-1}$ for which the exact solution is known. In 
this case we obtained $\sigma=(\pi/\sqrt{3})a$ and 
$\mu_2^3=(3\sqrt{3}/2\pi^3)\zeta(3)\approxeq 0.100$.  Assuming 
further $\mu_2^3=0.1$ the model X-ray profile corresponding to the
Al$_2$O$_3$/Cr(70\AA)/[Fe(20\AA)/Cr(9\AA)]$_8$
multilayer have been calculated, taking into account the possible asymmetry 
$\mu_1$ in the interfacial structure. Provided the matrices $S_{k,k+1}$ are 
known, the solution of the Eq.~(\ref{Helm}) and, hence, the scattering matrix S 
of the whole multilayer is found by means of recurrent scheme~\cite{Parrat}. The 
results obtained are shown in Fig 2. In agreement with Eqs.~(\ref{r11}-
\ref{t12}) the phase correction becomes essential with the increase of the 
incident angle $\theta$ and it provides a more adequate description of the 
reflectometry spectrum for the scattering vectors in the range from the first 
to the second Braggs' peaks. The another sample profile is shown in 
Figs.~3 and 4.  It corresponds to the structure
Cr$_2$O$_3$($\sigma$=3\AA)/Fe($\sigma$=2\AA)/[Cr($\sigma$=2\AA)/Fe($\sigma$=2\AA)]$_{11}$/Cr
($\sigma$=3\AA)/Al$_2$O$_3$($\sigma$=1\AA); 
Cr$_2$O$_3$($\mu_1$=0.77)/Fe($\mu_1$=1.25)/
[Cr($\mu_1$=-0.0045)/Fe($\mu_1$=-1.0)]$_{11}$/Cr($\mu_1$=-0.56)/Al$_2$O$_3$
($\mu_1$=2.4). 
The more exhaustive account and the details of our 
numerical algorithm will be presented elsewhere. We would like to emphasize 
that the form of the scattering matrix as given in Eqs.~(\ref{r11}-\ref{t12})  
is rather general, i.e., it is irrelevant to the precise form of a reflectivity 
profile. Thus it provides the unification description of a large variety of 
possible symmetric as well as asymmetric interfaces.

Summing up, we have developed the theory of specular X-ray reflectivity 
from a rough interface based upon the reflection function method. By using the 
approximation of the abruptly changing potential we have found the phase 
correction to the reflectivity due to interface roughness and asymmetry, which 
is essential for the description of the X-ray reflectivity spectra for greater 
incident angles.
\vspace{0.5cm}

The research was partially supported by RFBR (Grants No. 01-02-17119 and 
00-15-96745 ).


\end{document}